\documentclass[twocolumn,showpacs,preprintnumbers,amsmath,amssymb,floatfix]{revtex4}

\usepackage{graphicx}

\newcommand{\idest}{{\it i.e.\/}}
\newcommand{\etal}{{\it et al.\/}}
\newcommand{\Journal}[4]{#1 {\bf #2}, #3 (#4)}

\newcommand{\oneu}{$1_u$}
\newcommand{\twou}{$2_u$}
\newcommand{\zeroup}{$0_u^+$}
\newcommand{\singS}{$2^1$S}
\newcommand{\tripS}{$2^3$S$_1$}
\newcommand{\tripP}{$2^3$P$_2$}
\newcommand{\quintet}{$^5\Sigma_g^+$}
\newcommand{\triplet}{$^3\Sigma_u^+$}
\newcommand{\singlet}{$^1\Sigma_g^+$}
\newcommand{\quintetP}{$^5\Sigma_u^+$}

\newcommand{\singletP}{$^1\Sigma_u^+$}
\newcommand{\ee}[1]{10$^{#1}$}
\newcommand{\EE}[2]{#1$\times10^{#2}$}
\newcommand{\anul}{a$_0$}
\newcommand{\mat}[1]{\mathbf{#1}}

\begin{document}

\title{On the Role of Penning Ionization in Photoassociation Spectroscopy}

\author{E. van der Zwan}
\author{D. van Oosten}
\author{D. Nehari}
\author{P. van der Straten}\email{P.vanderStraten@phys.uu.nl}
\affiliation{Atom Optics and Ultrafast Dynamics, Utrecht University,\\
P.O. Box 80.000, 3508 TA Utrecht, The Netherlands}

\author{H.T.C. Stoof}
\affiliation{Institute for Theoretical Physics, Utrecht University,\\
P.O. Box 80.195, 3508 TD Utrecht, The Netherlands}

\date{\today }

\begin{abstract} 
We study the role of Penning ionization on the photoassociation spectra of He(\tripS)-He(\tripS). The experimental setup is discussed and experimental results for different intensities of the probe laser are shown. For modelling the experimental results we consider coupled-channel calculations of the crossing of the ground state with the excited state at the Condon point. The coupled-channel calculations are first applied to model systems, where we  consider two coupled channels without ionization,  two coupled channels with ionization, and  three coupled channels, for which only one of the excited states is ionizing. Finally, coupled-channel calculations are applied to   photoassociation of He(\tripS)-He(\tripS) and good agreement is obtained between the model and the experimental results.
\end{abstract}

\maketitle

\section{Introduction}

Photoassociation spectroscopy (PAS)  is a powerful technique for the study of ultracold collisions. In the case of alkali-metal atoms, the technique has been used to study long-range molecular states,  the scattering length of the interacting atoms, the lifetime of the resonant state, and the dynamics of the interacting system~\cite{pas}. In PAS two interacting ground-state atoms absorb a photon from the laser beam and are excited to an excited state of the molecule. The detection of the PAS process is in many cases performed by observation of trap loss, where one of the interacting atoms gains sufficient energy in the process to be lost from the trap. Since the relative energy at low temperatures is small, PAS is able to produce very high resolution spectra ($<$1 MHz). This allows for the detailed study of singly excited molecular states, where the interaction takes place at long range close to the dissociation limit. Since the interactions at these distance are caused by dispersion forces, which are determined mainly by atomic parameters, detailed comparison between theory and experiment is possible for such systems. Since the strength of the PAS resonances depends on the overlap between the ground and the singly excited state at the Condon point, the progression of the intensities of PAS lines allows for a determination of the wavefunction of the ground state and thus for a determination of the scattering length, which is important for the achievement of Bose-Einstein condensation. Due to this feature PAS has become a proven technique for interacting alkali-metal atoms.

In the case of metastable rare-gas atoms, the situation is different. Two interacting,  metastable rare-gas atoms have sufficient energy for the ionization of one of the atoms leading to the well-known process of Penning ionization. The excitation of singly excited states in such systems can be accompanied by a strong increase of Penning ionization, which may prohibit the observation of narrow resonances. Although the detection of ions and/or electrons can be much more sensitive than the detection of fluorescense as a measure of trap loss, observation of PAS in metastable rare-gas systems using ion detection might not be the appropriate method. However, PAS has been observed a couple of years ago using ion detection for interacting, metastable helium atoms~\cite{iondetection}. Since that time  studies have been carried out for this system using both ion and fluorescense detection~\cite{helium1,helium2,helium3}. Recently, the results of two groups for PAS on metastable helium atoms have been compared, where one group used ion detection and the other group applied trap loss detection~\cite{ENSandAOpaper}. The study showed, that the results of these two different techniques agreed within experimental resolution. This is especially remarkable, since both the temperature and the densities between the two experiments  differed by three orders of magnitude.

In this paper we study the role of Penning ionization on the PAS process. After discussing the technique of PAS for metastable helium and the results of our experiment, we introduce a pedagogical model for the PAS process, where the ground and excited channels are coupled by the light interaction. In this model we first consider the interaction of one ground-state channel with one excited state channel without ionization. We show how the light interaction leads to the observation of the resonances in the elastic scattering rate, which can be ascribed to bound states in the excited-state potential. We also show that by increasing the Franck-Condon overlap between the ground and the excited state, the resonances broaden and shift due to the interaction with light. If we allow the excited state to become ionizing, the bound states obtain a finite lifetime and now also lead to resonances in the ionization cross section. The results of this two-channel problem are compared to a semi-classical Landau-Zener model. We find good agreement, if we optimize the parameters of the Landau-Zener model with respect to the coupled-channel calculations. In a next step we expand our model by including a second, excited state and assuming that only one of the excited states is ionizing. This leads to the observation of PAS resonances  of the non-ionizing channel through the coupling with the ionizing channel. Again, the results are compared with a Landau-Zener model. The model clearly shows how the Penning ionization process leads to the observation of the PAS resonances through ionization.

Finally we apply our multi-channel model to investigate PAS of the He(\tripS-\tripS) system. In Ref.~\cite{ENSandAOpaper} it has been shown that all observed resonances in this system can be attributed to the coupling of the ``ground'' state channel with four excited-state channels. We solve the multi-channel coupled equations for this system and obtain good agreement with the experimental results. This shows that the insights obtained from the simple, pedagogical model can be applied to this complicated system of many coupled channels. Although the results we show are specific for the metastable helium system, they can easily be generalized to the other metastable rare-gas systems and also to other systems, where ineleastic processes play an important role.

\section{Experimental setup}

The experimental setup consists of two main parts, the metastable helium source and the magneto-optical-trap (MOT) chamber, which are separated by three differential pumping stages. Metastable He(\singS) and He(\tripS) atoms are produced in a DC discharge by electron impact~\cite{woestenenk}. A Pyrex glass tube with an orifice of 1 mm inserted in a Teflon holder is placed in a copper assembly. The glass tube and nozzle assembly are placed in a cryostat. The gas flows on the outside of the glass tube along the cold cryostat and cold nozzle, before it expands into the vacuum chamber. Part of the gas is pumped out through the glass tube, as in the liquid nitrogen cooled DC-discharge source described by Kawanaka~\etal~\cite{kawanaka}.  However, in contrast to that source, in our design the source is cooled to a much lower temperature of about 10-15 K by liquid helium. The nozzle is made out of an aluminum plate of 1 mm thickness to reduce heat gradients (a copper nozzle did not perform as well as an aluminum one) and the exit hole diameter used is 0.5 mm. As a discharge needle we use a tungsten rod with a diameter of 2 mm, which is sharpened at the end and kept in the middle of the glass tube by ceramic spacers. The optimal distance from the nozzle plate to the discharge needle is found to be 7 mm, ensuring both a high output stability and a high yield of the source. To minimize heating effects the source is operated at a low discharge power of typically 35 mW (about 0.05 mA) and a low discharge pressure of \ee{-2} mbar. The helium discharge is localized inside the source, \idest, between the discharge needle and the nozzle plate. This is in contrast to conventional metastable sources, which operate at higher pressures (a few mbar) with a discharge burning outside the nozzle, \idest, between the discharge needle and the skimmer, to reduce quenching of the metastable atoms~\cite{fahey,ohno}. We observe that operation in this manner is only possible at higher source pressures and currents, and therefore at higher temperatures. However, we estimated the quenching of the metastables in our design to be of minor importance due to our low source pressure of \ee{-2} mbar. Operating at minimum power and pressure yields a total flux of a \EE{5}{12} atoms/s sr. The mean velocity of the atoms is only 300 m/s, compared to 2000 m/s in the case of conventional sources. Since the slowing distance for laser cooling is proportional to the square of the initial velocity, this leads to a reduction in slowing distance of nearly a factor 50~\cite{woestenenk}.

Therefore, the slow atoms produced can be loaded directly in a magneto-optical trap (MOT) without the need of a Zeeman slower. Usually a He(\tripS) MOT is loaded from a discharge source cooled with liquid nitrogen~\cite{nitrogen1,nitrogen2,nitrogen3}.  In that case, the atoms need to be slowed in a Zeeman slower down from 1000 m/s to typically 60 m/s, before they can be captured in the MOT. Using a Zeeman slower makes the experiment more complex, since special care has to be taken to avoid reduction of density of the beam at the end of the slower~\cite{density}. In a Zeeman slower the beam is slowed down in the longitudinal direction and as it is slowed down, the transverse spreading of the beam becomes more dominant, thereby decreasing the flux in the center of the beam. This effect limits the number of atoms that can be captured in the MOT. This can in principle be compensated for by inserting a transverse cooling section at the end of the slowing unit, but this will complicate the experimental setup further.

To slow the atoms from the source at 300 m/s to the capture velocity of the MOT ($\approx$ 30 m/s), we slow the atoms in the MOT chamber. The coils that generate the MOT magnetic field in our setup are large in diameter (about 400 mm). These large coils produce a magnetic field that is sufficient for Zeeman slowing and thus no additional Zeeman slower is needed. The atoms are slowed by a counterpropagating slowing laser detuned 50 MHz below the He(\tripS - \tripP) transition with a saturation parameter of $s_0$ = 400. The beam still fans out in the slowing process, but the effect is smaller than for a liquid nitrogen cooled source for two reasons. First, due to the lower initial velocity of the atoms the beam does not spread out as much during the deceleration. Second, since the atoms are slowed down in the MOT chamber the beam spreads out mostly in the region, where the MOT trapping laser beams are present. This reduces the losses considerably. The increased loading efficiency fully compensates for the low flux of the source and we find that the loading rate of the MOT from this source and the liquid nitrogen source~\cite{nitrogen1} are comparable.

The MOT laser beam is expanded by a telescope to a beam with radius of 10 mm and the saturation parameter of the laser light after the telescope is $s_0$ = 180. The laser frequency is locked by using the saturated absorption spectroscopy 14 MHz below the He(\tripS - \tripP) transition. We trap about  \EE{3}{5} atoms, as measured by releasing the atoms onto  micro-channel plates (MCP) located 70 mm from the center of the MOT chamber. The average density of the trapped atoms is \ee{9} atoms/cm$^3$ and the temperature is 1.9 mK, as measured by time-of-flight of atoms from the MOT, using the MCP.

\begin{figure}
    \centerline{\includegraphics[width=0.45\textwidth]{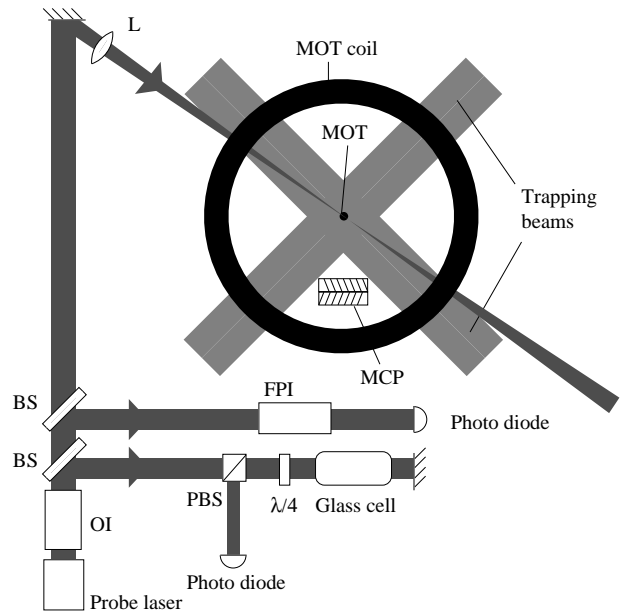}}
    \caption{Setup of the probe laser. The MOT is overlapped with the probe laser, which is nearly collinear with one of the MOT laser beams. A small fraction of the probe laser light is split off for the Fabry-P\'erot interferometer (FPI) for relative frequency calibration, and for the saturation spectroscopy setup for a frequency reference with respect to the He(\tripS - \tripP) transition. The ions produced are detected with multi-channel plates (MCP).   \label{fg:probesetup}}
\end{figure}

An overview of the probe laser setup is shown in Fig.~\ref{fg:probesetup}. The probe laser beam is overlapped with the MOT cloud. To obtain a high probe light intensity, the probe beam can be focused by a lens. Additionally, a one-to-one telescope is used to overlap the focus of the probe beam with the MOT cloud. In order to calibrate the frequency scale, two beam splitters each split off 5\% of the light from the main laser beam. One beam runs through a Fabry-P\'erot interferometer (FPI) with a free spectral range of $\approx$ 1 GHz and the transmitted signal is detected by a photo diode. This is used as a  relative frequency reference. The other laser beam runs through a saturated absorption spectroscopy setup, which is used to calibrate the offset of a frequency scan with respect to the He(\tripS - \tripP) transition~\cite{Michiel}.

In the experiment, pairs of cold atoms can absorb a photon from the probe laser beam and be excited to an long-range attractive potential. The probe laser is scanned in frequency below the D$_2$ line and the ionization signal caused by Penning collisions in the MOT is studied as a function of the probe laser frequency. The peaks observed in the ion rate are attributed to vibrational states of the excited molecules. Close to the atomic D$_2$ resonance we observe a large dip in the ionization spectrum, which corresponds to the fact that close to resonance atoms are heated due to the radiation pressure of the probe laser, in which case no atoms could be trapped or photoassociated in the MOT.

To reduce the very high count rate of Penning ionization due to the MOT lasers, we periodically modulate the MOT laser (with a frequency 25 kHz) with a square wave modulation, such that during the ``off'' period the MOT laser is detuned by 500  MHz below the resonance frequency. This increases the signal-to-noise ratio considerably. The spectra are recorded during the trap ``off'' period and in general 20 spectra are averaged.

\begin{figure}
    \centerline{\includegraphics[width=0.45\textwidth]{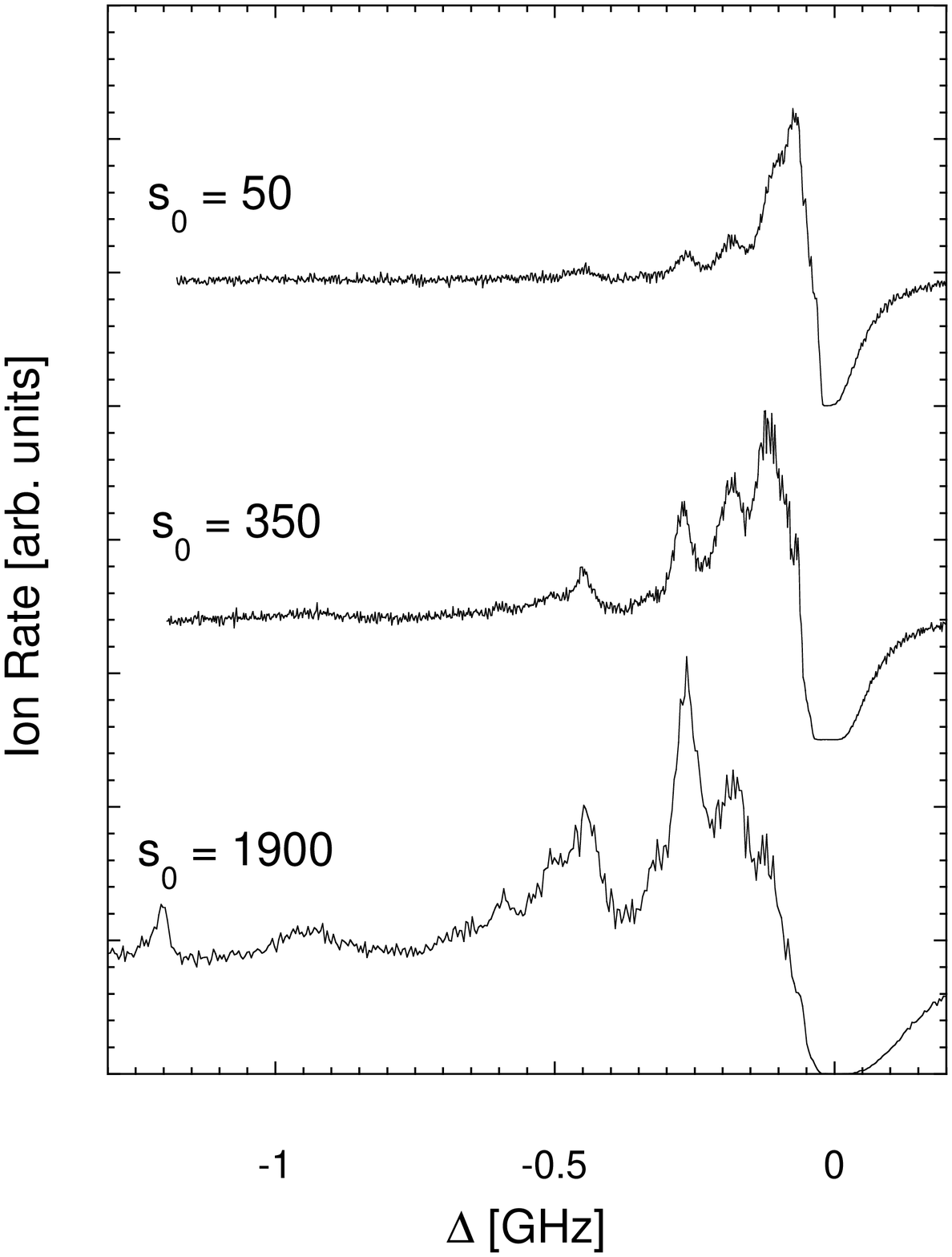}} \strut\\
    \centerline{\includegraphics[width=0.45\textwidth]{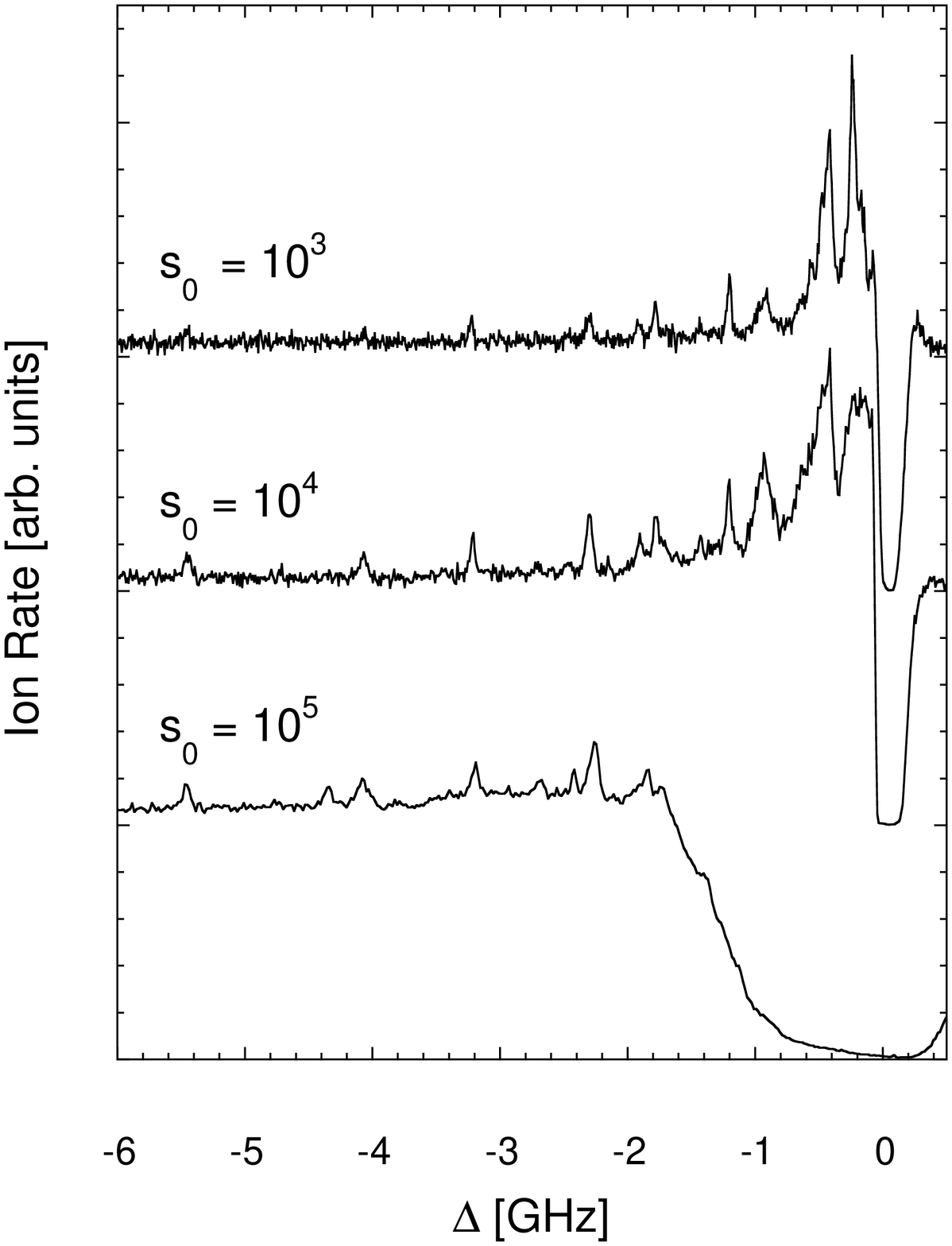}}
    \caption{ Spectra measured with increasing probe laser intensity ranging from $s_0$ = 50--\ee{5}. Note, the change in  detuning scale between the two panels.  \label{fg:expspec}}
\end{figure}

We have measured photoassociation spectra for various probe laser intensities. In Fig.~\ref{fg:expspec} a number of spectra are shown which have been measured with increasing probe laser intensities. The intensities range from $s_0$ = 50--\ee{5}. The figure shows that, when the laser intensity increases, ionization peaks start to appear at increasing detunings and that the heights of the peaks become larger. Deeply bound states can only be measured with high probe laser intensities. The bound states that lie close to the dissociation limit are measured with low intensities since a high intensity probe laser would disturb the cloud of atoms in the frequency region of interest. We have measured peaks at detunings down to $\approx$ 14 GHz below the D$_2$ line. 

\section{Theoretical model}

\subsection{Ground molecular state}

In the experiment, atoms are prepared in the He(\tripS) state. This state is metastable with a lifetime of 8000~s and thus  stable on the time-scale of the experiment. We  therefore consider it to be the ``ground'' state. Two atoms colliding in the ``ground'' state  interact through the \singlet, \triplet\ or \quintet\ potential. We  only consider in our analysis the \quintet\ state based on the fact that its multiplicity is much higher than that of the \singlet\ state, and that the \triplet\ state is repulsive. Because of the low temperature we also only consider $s$-wave scattering ($l=0$).

In our calculations we use the long-range \quintet\ potential as calculated by St\"{a}rck~\etal~\cite{Starck}, given by the following analytical fit of their numerical calculations (atomic units are used in the description of the theoretical model)
\begin{equation}
V(r)=(c_1+c_2r^{c_3})e^{-c_4r-c_5r^2}+\sum_{n=3}^{p}\frac{C_{2n}\Gamma_{2n}
[c_6(r-c_7)]}{r^{2n}} ,
\label{groundstatepotential}
\end{equation}
where
\begin{equation}
\Gamma_{2n}(x)=1-\left[ e^{-x}\left( \sum_{m=0}^{2n}\frac{x^m}{m!} \right) \right] ,
\end{equation}
and the constants are given by $c_1$ = 0.247643, $c_2$ = \EE{6.9178}{-6}, $c_3$ = 5.851244, $c_4$ = $-$0.262988, $c_5$ = 0.166841, $c_6$ = 0.770951, $c_7$ = 1.335894, $C_6$ = \EE{3.265}{3}, $C_8$ = \EE{2.075}{5}, $C_{10}$ = \EE{2.107}{7}, and \mbox{$C_{2n>10}=C_{2n-6}(C_{2n-2}/C_{2n-4})^3$}. In their paper the authors state that they use $p=5$. However, we obtained the best fit to their data if we allowed the sum to go up to $p=8$.

\subsection{Excited molecular states}

\begin{figure}
    \centerline{\includegraphics[width=0.45\textwidth]{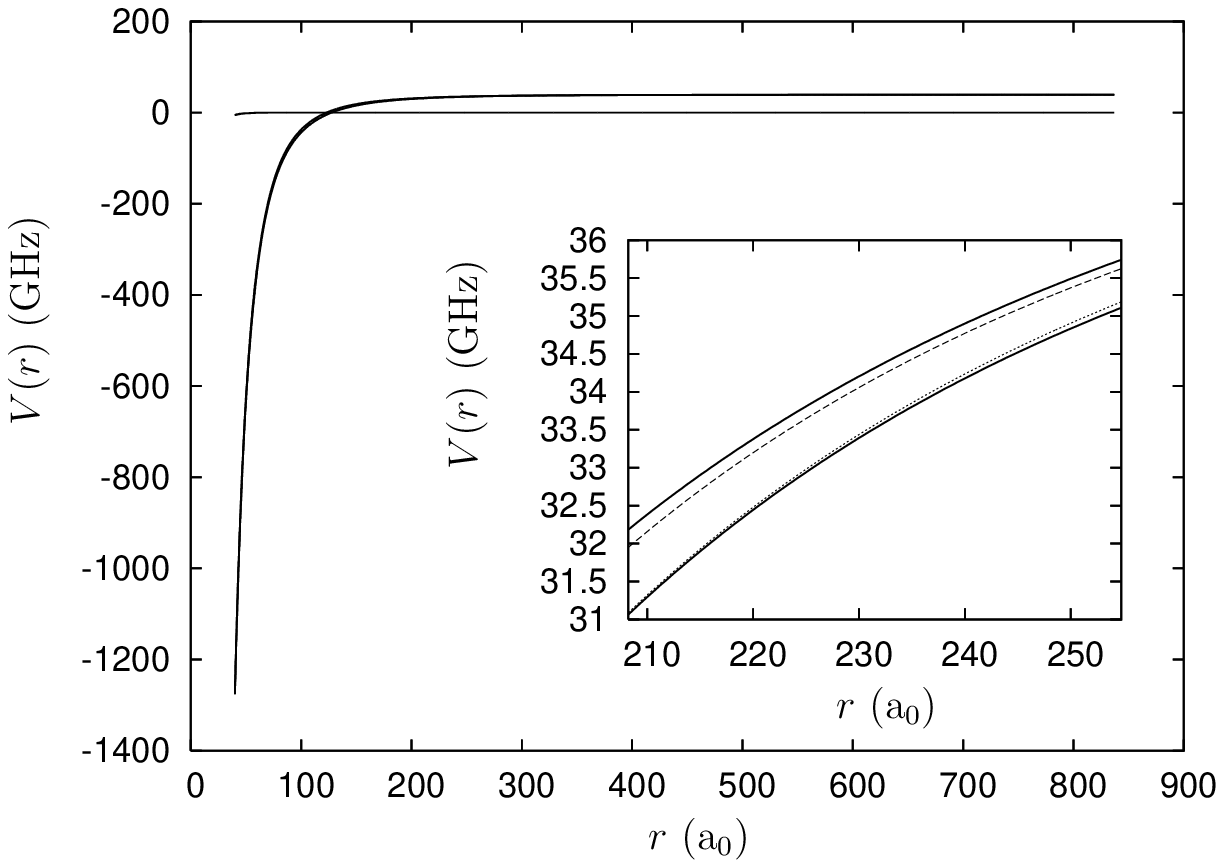}}
    \caption{The ground and four excited-state potentials  \zeroup\ (solid lines), \oneu\ (dashed line) and \twou\ (dotted line). The excited-state potentials are shifted down by the \tripS-\tripP\ resonance frequency.} \label{heliumpotentials}
\end{figure}

The probe laser is detuned below the He(\tripS-\tripP) transition. There are many potentials that are connected to this asymptote, which have previously been calculated by Mastwijk~\etal~\cite{nitrogen1} and more recently by other groups~\cite{venturi,Dickinson}. In a joint paper by the ENS group and our group~\cite{ENSandAOpaper} it has been shown that only four of these potentials are relevant for the photoassociation, indicated by \zeroup\ (twice), \oneu\ and \twou\ (Hund's case (c) notation). In Figure~\ref{heliumpotentials} the five potentials used in the model are shown. The \zeroup\ potentials are connected at short range with the \singletP\ potential, which ionizes with a probability $p_i$ of $\sim$ 80\%. The other two states connect at short range to the \quintetP\ potential, which is not ionizing. The splitting between these potentials is due to the fine structure coupling.  The couplings between these excited-state potentials were calculated {\it a priori} and included in the model. It should be noted that there is no direct coupling between the ionizing \zeroup\ states and the \twou\ state. That means that the \twou\ channel can only couple to the ionizing channels via the \oneu\ channel.

\subsection{Extended Numerov method}

The individual channels in our system without coupling are described by the radial Schr\"{o}dinger equation, given by
\begin{equation}
\frac{d^2\psi(r)}{dr^2}+k^2(r)\psi(r)=0,
\label{radialSchrodingerEquation}
\end{equation}
where $ k^2(r)\equiv 2\mu[E-V(r)]$ and $V(r)$ is the interatomic potential. Furthermore, $E$ is the relative kinetic energy of the atoms and $\mu$ is the reduced mass of the system. To perform the numerical calculations presented in this article, we use an extended version of the Numerov method to solve the radial Schr\"{o}dinger equation~\cite{johnson}. In its original form, the Numerov equation is a discretized version of the one-channel Schr\"{o}dinger equation. The integration domain is divided into $N$ steps of size $h$, such that $r_n=r_0+n h$. In that case, the Numerov equation can be written in a two-point form, which can be numerically integrated faster~\cite{johnson}. In the multi-channel calculations the wavefunction is written in matrix form, where each row represents an asymptotic eigenstate of the system (channel) and each column a linearly independent solution. The Hamiltonian contains off-diagonal elements that describe the couplings between these channels. For a model system consisting of a ground state ($s$) and two excited-state channels ($p_1$, $p_2$) with a $r$-independent rotational coupling $J$ between the excited states, the difference equation becomes
\begin{equation}
 \mat{F}_{n+1} = (2\mat{I}+10\mat{T}_n)\mat{\psi}_n-\mat{F}_{n-1},
 \label{NumerovEquation}
\end{equation}
with
\begin{equation}
\mat{T}_n = -\frac{1}{12}h^2
\begin{pmatrix}
k_{s}{}^2(r_n) & \mu\Omega & \mu\Omega\\
\mu\Omega & k_{p_1}{}^2(r_n) & J \\
\mu\Omega & J & k_{p_2}{}^2(r_n)
\end{pmatrix}.
\end{equation}
Here $\Omega$ is the Rabi coupling between the ground and excited states, $\mat{I}$ is the identity matrix, $\mat{F}_n\equiv (\mat{I}-\mat{T}_n)\mat{\psi}_n$ and $k_{i}(r_n)$ is the wavevector of state $i$ at $r_n$. In this equation the wavefunction occurs explicitly, so inversion of $(\mat{I}-\mat{T}_n)$ is needed at every integration step.

\begin{figure}
    \center{\includegraphics[width=0.45\textwidth]{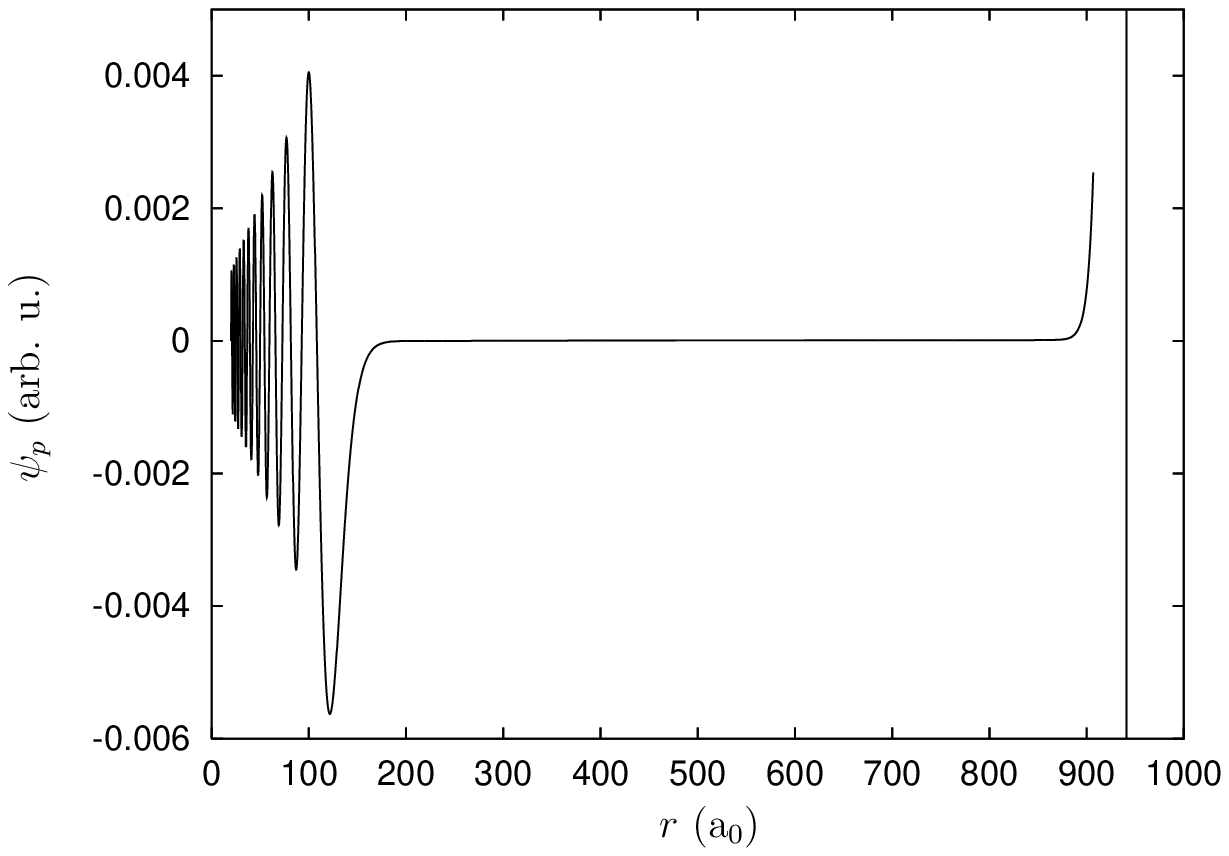}}
    \caption{Excited-state wavefunction.
    The vertical line indicates the last point where we applied the
    diagonalization procedure.} \label{PP-wave2}
\end{figure}

\hyphenation{dia-go-na-li-zation} 
\hyphenation{ana-ly-tically} 
We start integrating outwards with some initial condition of the components of the wavefunction of the closed channels. Once we pass the Condon point an excited-state channel becomes closed. Numerically integrating the wavefunction in a closed channel gives rise to two solutions; a physical solution, that is exponentially decaying, and a non-physical solution, that is exponentially growing. We exploit a diagonalization procedure~\cite{Stoof} to transform away the non-physical solution. The procedure works as follows: At a point $r_m$, where the absolute value of one component of $\mat{F}_m$ reaches a critical value that is much larger than unity (for example 10), which indicates that the wavefunction has grown significantly and therefore that the non-physical solution has become dominant in that channel, the inverse of $\mat{F}_m$ is calculated and applied to all $\mat{F}_n$ with $n\leq m$. This selects the initial conditions such that after integration $\mat{F}_m$ becomes equal to the identity matrix. With respect to before the inverse of $\mat{F}_m$ is applied, the norm of the wavefunction at $r_m$ has decreased significantly. In fact, it is now such that we can conclude that the non-physical solution has not become dominant at $r_m$. Since the non-physical is exponentially growing, this means that it does not have a significant amplitude at a short distance before $r_m$. So effectively we have transformed away the non-physical solution to select the ``right'' physical initial conditions. After continuing with the integration, we apply the diagonalization procedure again whenever the critical value is reached. In this way we tune the initial conditions such that the non-physical solution does not become dominant for increasing $r$. The result of this procedure can be seen in Fig.~\ref{PP-wave2}, which is an example for an excited-state wavefunction. From the figure it is clear that the wavefunction is physical up to some small distance before the last point where the diagonalization procedure is applied. 

\subsection{Accumulated phases}

For metastable helium the molecular potentials in the short-range domain ($r<$20\anul) are not known accurately enough to use them for our integration procedure. Therefore we employ the \emph{accumulated phase method}~\cite{Verhaar}. In this method the effect of the short-range potential is replaced by a pre-determined phase, which represents  the phase accumulated over the short-range region. The value of this accumulated phase will depend on the details of the inner region of the potential, which are unknown. However, if the potential is sufficiently deep compared to the energy of the bound states, this phase is independent of the exact position of the bound states and thus one accumulated phase is sufficient to calculate all bound states close to the dissociation limit. As a result, the accumulated phases can  be fitted to match the experimental results. In the system under study some of the channels are ionizing in the inner region through Penning ionization. We take this into account by taking the accumulated phase to be a complex quantity in this case, where the real part is the actual phase shift of the wavefunction and the imaginary part takes into account the ionization.

\subsection{Integration}

For the numerical integration of the wavefunction, the overall error in the integration is important. Blatt~\cite{Blatt} derived that for the Numerov method the relative error
per step $\varepsilon_1$ can be approximated by
\begin{equation}
\varepsilon_1 = \frac{\mbox{Error in } \psi_{n+1}}{\psi_n} \approx - \frac{72}{10} T_n^3 \propto  h^6 k^6\sim h^6 [E-V(r)]^3.
\end{equation}
However, as shown by Sloan~\cite{Sloan} the error $\varepsilon_N$ after many steps becomes proportional to $h^4$. We find that
\begin{equation}
\varepsilon_N \sim h^4 \left|E-V(r)\right|^3.
\label{totalerrorEquation}
\end{equation}
This shows that it will be advantageous to have our stepsize dependent on $|E-V(r)|$ with a smaller stepsize for the region where $V(r)$ is still very deep. This is the basis for the Fourier grid method~\cite{fourier}. However, in our case, where we are dealing with a few coupled equations, we restrict ourselves to changing the stepsize $h_1=h$ once to a new stepsize $h_2=k h_1$, where $k$ is a positive integer.

In practice we use 100.000 steps for the one-channel calculations to find the bound states in the potentials. The integration range runs from 20 to 900\anul. Half the number of steps are used for the first 80\anul, whereas the other half of the steps are spread out over the remaining 800\anul. For the multi-channel calculations we integrate from 20 to 2000\anul\ using 10.000 points. Half these points are equally spread between 20 and 60\anul, and the other half is spread out over the rest of the integration range. Depending on the detuning, the Condon points for these four potentials generally lie between 120 and 240\anul\ in our calculations. Also depending on the detuning, we need to apply the diagonalization procedure somewhere between 10 and 100 times. The total flux is then conserved over the integration range to an accuracy of more than 98\% for the final calculations presented in this paper. This is an indication of the accuracy of the procedure.

\subsection{Extracting cross sections}

The key to extracting cross sections from the calculated wavefunctions is the phase shift $\delta_\ell$ in the open channel (ground state). This phase shift describes the asymptotic effect of the collision between the atoms. Asymptotically the component of the wavefunction in the open channel can be written as~\cite{Joachain}
\begin{equation}
R_\ell(k,r) = A_\ell(k) \left( j_\ell(k r) - \tan \delta_\ell(k) \; \eta_\ell(k r) \right),
\label{exteriorsolution}
\end{equation}
where $\ell$ is the partial wave, $j_\ell$ and $\eta_\ell$ are the spherical Bessel and Neumann functions, respectively, and the $A_\ell(k)$ are prefactors. As mentioned earlier, in our numerical calculations we only consider $\ell=0$. In the asymptotic region the solution for the open channel reduces to
\begin{equation}
\psi_s(r)=A(k_s)\cos \delta_0 \; \sin(k_s r+\delta_0),
\end{equation}
where $k_s$ is the wavenumber for the $s$-channel in the asymptotic region, where it is no longer dependent on $r$. The logarithmic derivative of this last equation is given by
\begin{equation}
\gamma(r)=\frac{1}{\psi_s(r)} \frac{\partial \psi_s(r)}{\partial r} = \frac{k_s
\cos(k_s r) - k_s \tan \delta_0 \sin(k_s r)}{\sin(k_s r) - \tan \delta_0
\cos(k_s r)}.
\end{equation}
We can solve this for $\tan \delta_0$ to find~\cite{Joachain}
\begin{equation}
\tan \delta_0 =\frac{k_s\cos(k_s r)-\gamma(r)\sin(k_s r)}{k_s\sin(k_s r) +
\gamma(r) \cos(k_s r)}.
\label{tandeltaEquation}
\end{equation}
The logarithmic derivative in the Numerov method is given by~\cite{johnson}
\begin{equation}
\gamma_n \equiv \frac{\psi_n'}{\psi_n} = \frac{1}{h} \left( A_{n+1}R_n - \frac{A_{n-1}}{R_{n-1}} \right) (1-T_n),
\label{logderivative}
\end{equation}
where $A_n\equiv{(\frac{1}{2}-T_n)}/{(1-T_n)}$. We now use these two equations to determine the phase shift $\delta_0$ in the open channel. The point where we determine the phase shift should lie a short distance before the point where we last applied the diagonalization procedure to ensure that the non-physical solution does not contribute to the results. Since we consider ionization in the photoassociation process, the atoms can scatter either elastically or inelastically, which in our case is ionization. Using the complex phase shift  $\delta_0=\delta_{0,R}+ i \delta_{0,I}$  we denote the elastic cross section $\sigma_{\rm elas}$ as~\cite{Joachain}
\begin{equation}
\sigma_{\rm elas} = \frac{4\pi}{k_s^2}\sin^2\delta_{0,R},
\end{equation}
and the inelastic cross section $\sigma_{\rm inelas}$ as
\begin{equation}
\sigma_{\rm inelas} = \frac{\pi}{k_s^2}\left[1-e^{-4\delta_{0,I}}\right].
\end{equation}
The last expression will be used for the identification of the ionization process in our experiment.

\section{Model calculations\label{sc:model}}

\subsection{Introduction}

\hyphenation{theo-re-ti-cal}
In this section we  examine in detail some simple model systems that illustrate the physics of the ionization process in the metastable helium experiment. A theoretical introduction is given before we describe the model systems. The model systems consist of a flat ground-state potential and one or two excited-state potentials. In both cases one excited-state potential is ionizing. Where applicable, we compare our numerical results to Landau-Zener type calculations.

\subsection{Avoided crossing}

\begin{figure}
\centerline{ \includegraphics[width=0.45\textwidth]{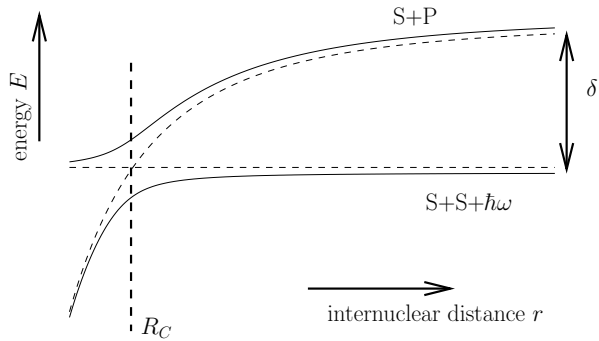}}
\caption{The uncoupled ``S+S'' and ``S+P'' molecular states (dashed lines) cross at the Condon point $R_C$  leading to an \emph{avoided crossing} of the coupled states (solid lines). \label{avoidedcrossingPicture}}
\end{figure}

The potentials in the helium system were calculated using the Born-Oppenheimer approximation and including the effects of rotational coupling between the different states~\cite{ENSandAOpaper,leonard}. The long range potential in the ``ground'' (``S+S'') state is a van der Waals $1/r^6$ potential and in the range of interest to us it can be considered flat. The excited-state (``S+P'') potential has a $1/r^3$ behaviour at long range.  For simplicity, we add the energy of one photon to the ground-state potential. For a negative detuning $\delta$ the shifted potentials cross at the Condon point $R_C$, as shown in Fig.~\ref{avoidedcrossingPicture}. Due to the coupling of the light field such a crossing becomes \emph{avoided} and the potentials in the vicinity of the crossing are modified. 

\subsection{Two channels without ionization}

In this model we neglect for simplicity the fine structure interaction in the \tripP\ state. Traditionally such a two-channel system can be approximately described  analytically using the Landau-Zener method. Using the approximations discussed above, we describe the potential for the ``S+P'' state as $V_{S+P}(r)=\delta+C_3/r^3$ with $C_3$ the dispersion coefficient for the dipole-dipole interaction. The Condon point is in this case given by $R_C=(-C_3/\delta)^{1/3}$. 

\begin{figure}
    \centerline{\includegraphics[width=0.45\textwidth]{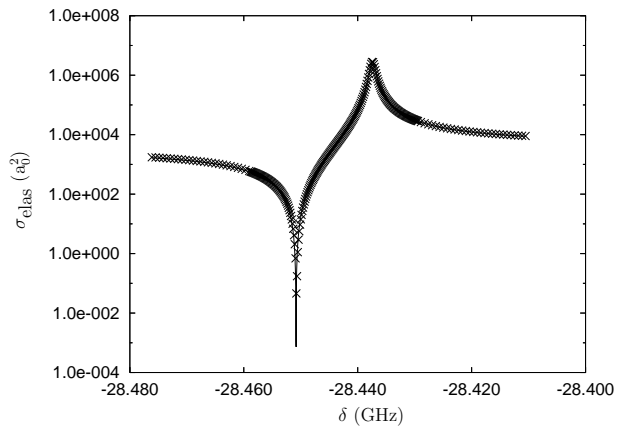}}
    \caption{Breit-Wigner profile in the elastic cross section (logarithmic scale) as a function of the detuning $\delta$ in the two-channel case with $C_3=-10$, $E = 190 \mu$K and $\Omega$ = 66 MHz. The crosses are numerical data and the line is a scaled Breit-Wigner profile with $E_r$ = $-$28.4374 GHz, $\Gamma$ = 998.8 kHz and $\phi$ = $-$\EE{3.721}{-2} rad. \label{2dbreitwigner}}
\end{figure}

An example of a photoassociation resonance is shown in Fig.~\ref{2dbreitwigner}, where we show the elastic cross section for the non-ionizing case. The elastic cross section consists of two parts. The first part is an almost constant background, which arises from the background scattering in the ground-state potential. The second contribution is a resonance when the energy of the incoming atom matches with a bound state in the excited-state potential. The two contributions interfere coherently to form a Breit-Wigner profile~\cite{Joachain} given by
\begin{equation}
 \sigma_{\rm elas} = \frac{\pi}{k_s^2} \left|\sin{\phi}\,
 e^{i\phi}+\frac{\Gamma/2}{(\delta-E-E_r)+i\Gamma/2}\right|^2,
 \label{BreitWigner}
\end{equation}
with $E_r$ the position of the resonance, $\Gamma$ the linewidth of the resonance and $\phi$ the phase difference between the direct scattering (the background) and the resonance.

\begin{figure}
    \includegraphics[width=0.4\textwidth]{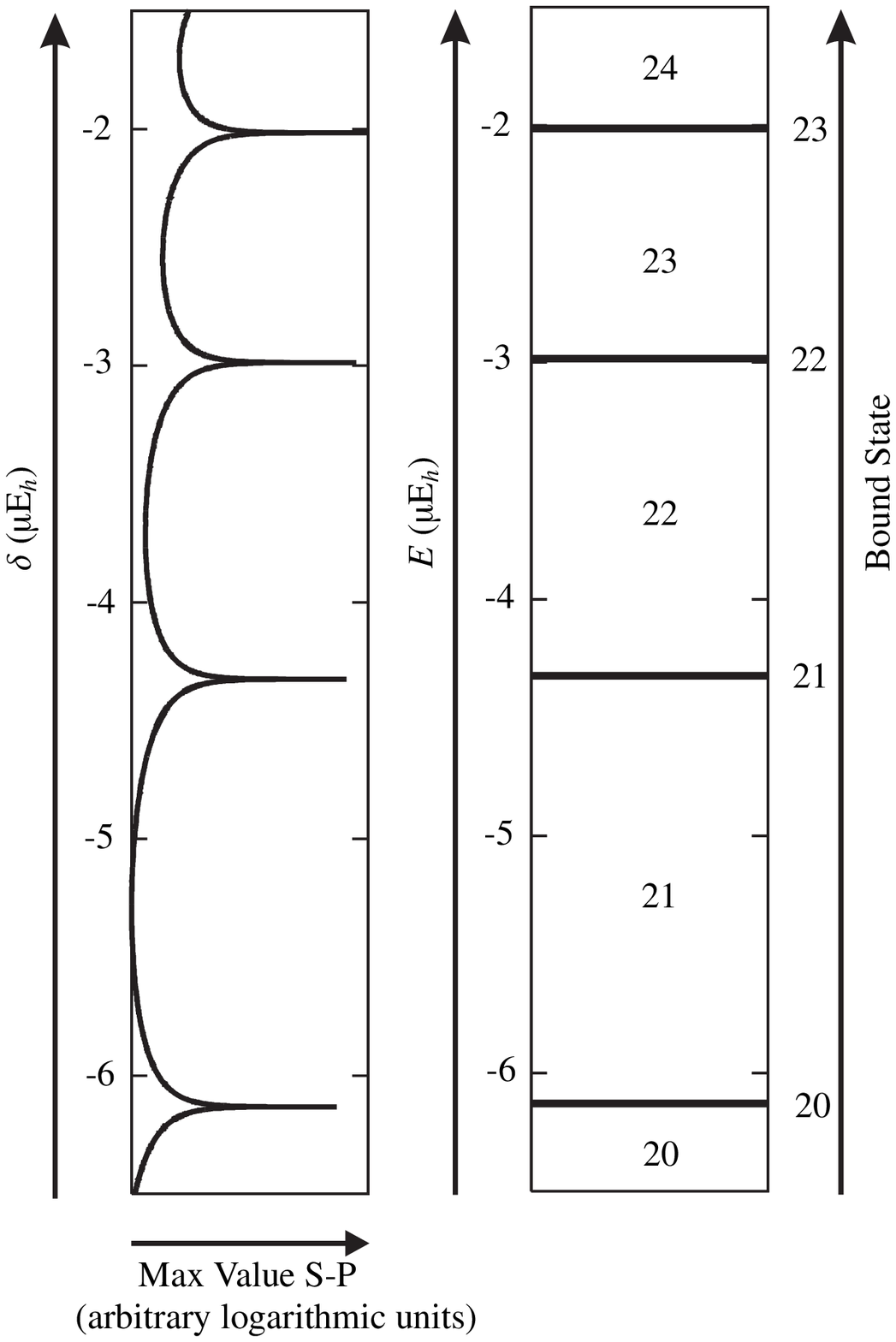} \centerline{(a)}
    \includegraphics[width=0.4\textwidth]{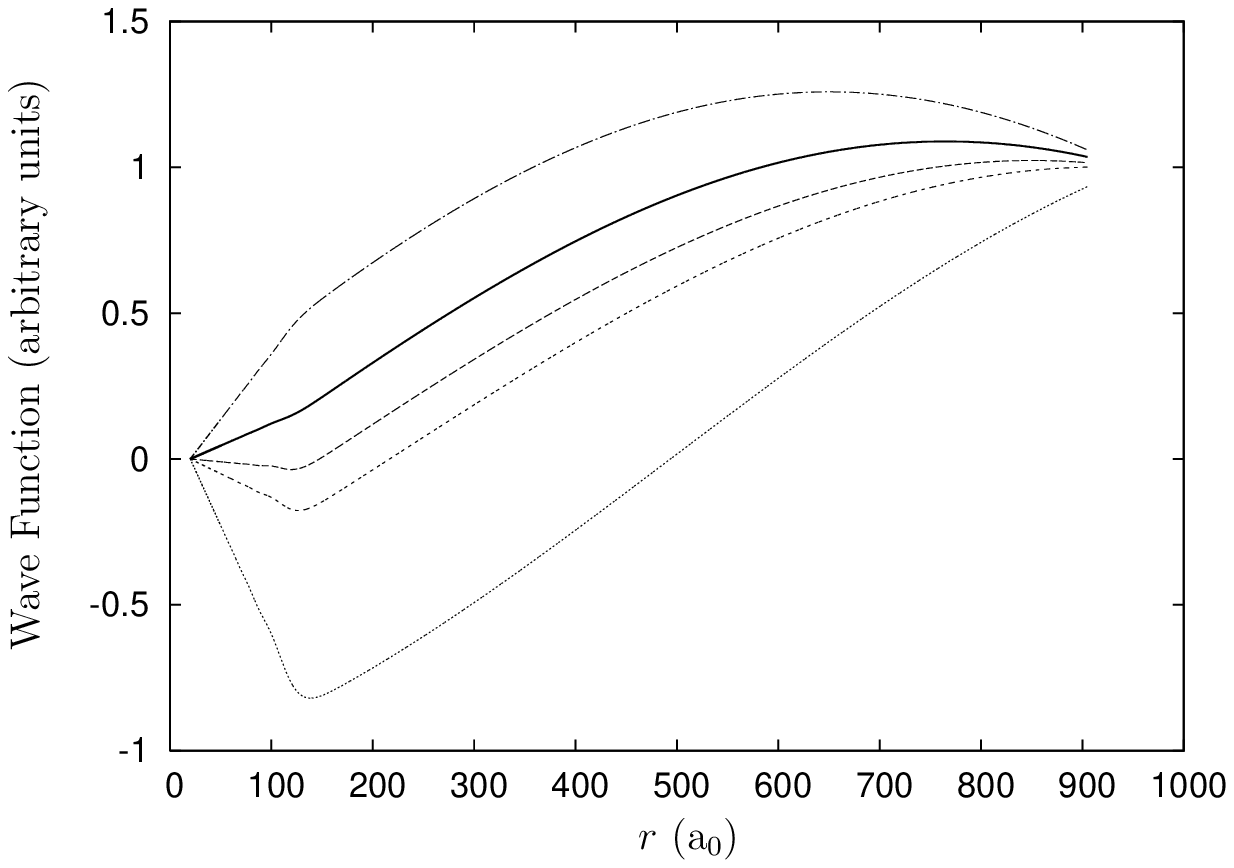}\\ (b)  \\
    \includegraphics[width=0.4\textwidth]{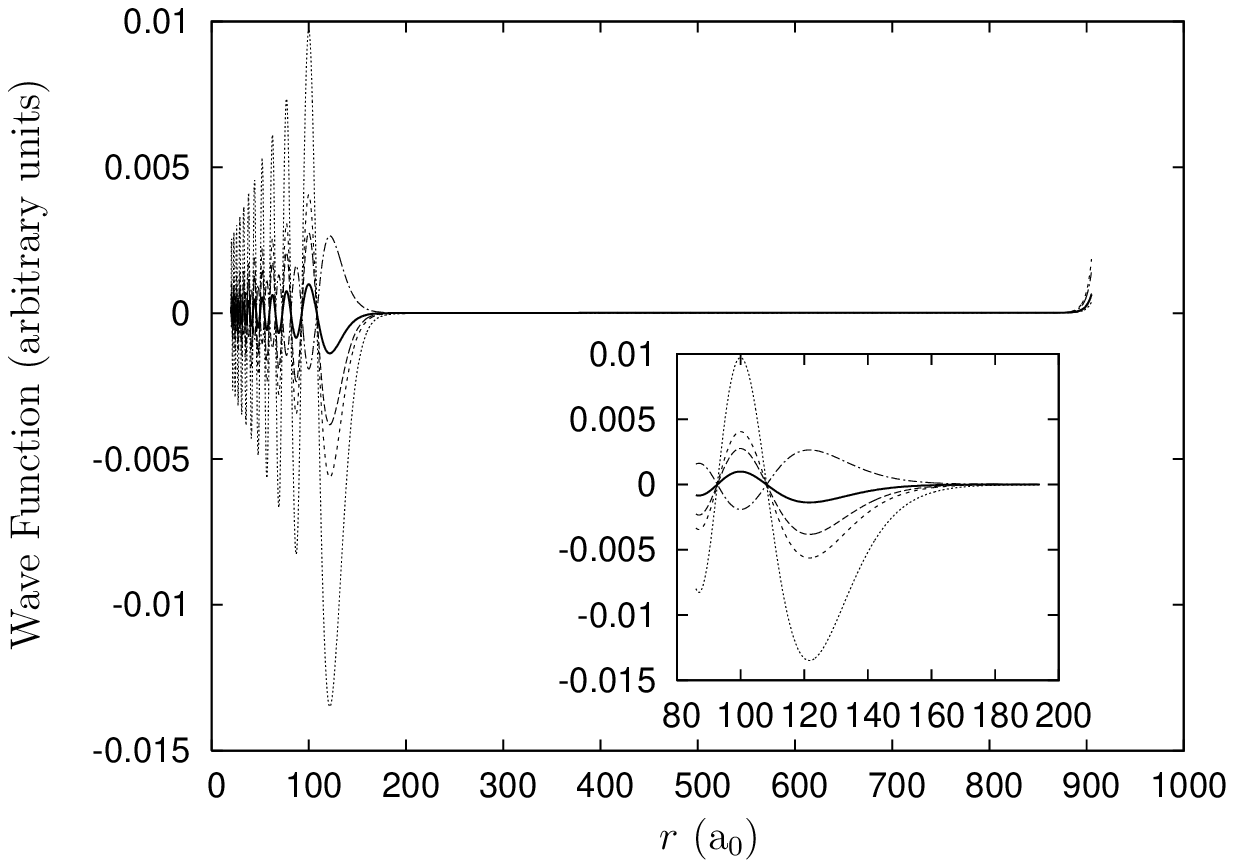}\\ (c)
    \caption{Right-hand side of Fig. (a) is an energy diagram. On the left-hand side a logarithmic plot of the maximum value the excited-state wavefunction reaches. Figure (b) is a plot of some ground-state wavefunctions around the $n=21$ resonance. The solid line is for $\delta$ slightly above the resonance, the dashed line is calculated at the resonance ($-$28.4374 GHz), the hashed line at the zero-crossing of the resonance, the dotted line at the lowest point of the resonance, and the dot-dashed line is for well below the resonance. Figure (c) is the same plot but now for the excited-state wavefunctions. \label{energydiagramc3=-10} }
\end{figure}

To take a closer look at what goes on at a photoassociation resonance, we consider Fig.~\ref{energydiagramc3=-10}a. On the right-hand side of this figure the bound states in part of the energy spectrum of the $C_3=-10$ potential for an accumulated phase of 0 rad at 20\anul\ are indicated. We assign numbers to these bound vibrational states by counting the nodes of the wavefunction. However, since we start to integrate at 20\anul, the number of nodes in the inner region is unknown. As a result a state numbered $n$ is not the $n^{th}$ vibrational state of the potential, but rather a state with number $n+m$, where $m$ is an offset. On the left-hand side of  Fig.~\ref{energydiagramc3=-10}a we show a plot of the maximum amplitude of the excited-state wavefunction in the two-channel calculation. It is clear from the graph that the wavefunction maximizes, whenever the energy corresponds to a bound state.

Some wavefunctions are drawn in Figs.~\ref{energydiagramc3=-10}b and \ref{energydiagramc3=-10}c. From Fig.~\ref{energydiagramc3=-10}b it is clear that the effective coupling for the three wavefunctions around the resonance is much stronger than for the other two, as their phases are altered much more dramatically. From Fig.~\ref{energydiagramc3=-10}c we can see that below the resonance (the dot-dashed line) we indeed have one node less than at resonance or above it. Also the artificial, exponential increase of the wavefunctions due to the renormalization procedure near the end of the integration range is visible.

\begin{figure}
\centerline{\includegraphics[width=0.45\textwidth]{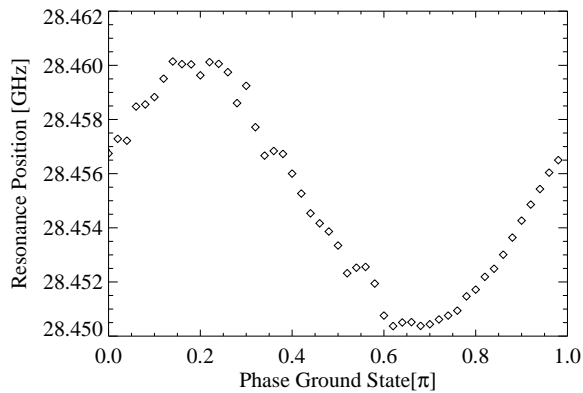}} (a) \\
\centerline{\includegraphics[width=0.45\textwidth]{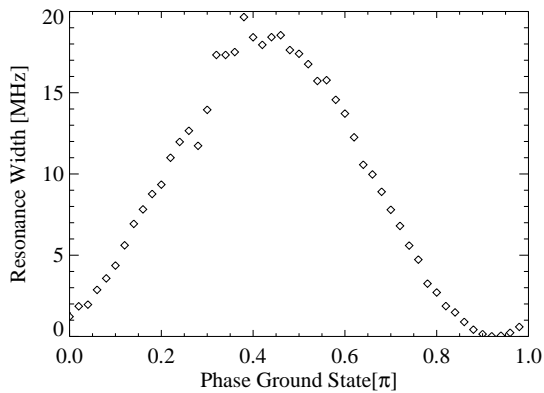}} (b) \\
\centerline{\includegraphics[width=0.45\textwidth]{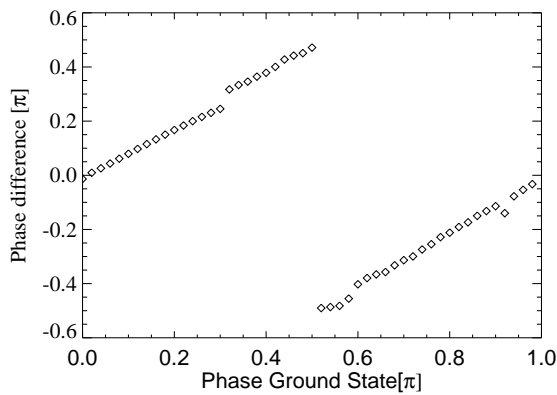}} (c) 
\caption{Resonance properties  as a function of the accumulated phase $\phi_s$ of the ``S+S'' channel at 20\anul: (a) Position of the resonance $E_r$, (b) width of the resonance $\Gamma$  and (c) phase difference $\phi$ between the direct scattering and the resonance. For small couplings the position of the resonance is at $-$28.43543 GHz. Due to the changing Franck-Condon overlap between the ground and excited state, both the shift and the width of the resonance change. \label{simplemodel}}
\end{figure}

\begin{figure}
\centerline{\includegraphics[width=0.45\textwidth]{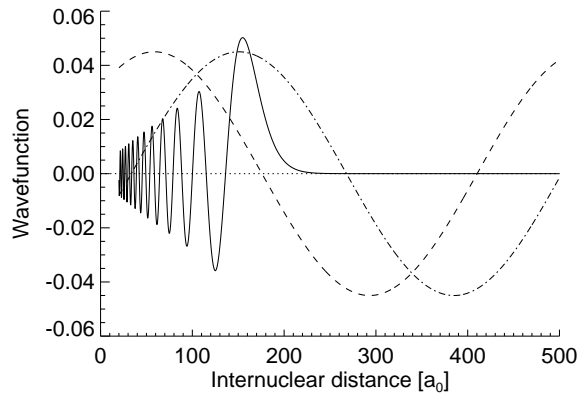}}
\caption{Wavefunctions of the ``S+P'' channel (solid line) and ``S+S'' wavefunction (dashed, dashed-dotted) for two different  accumulated phases of the ``S+S'' channel. For the first value of the accumulated phase (dashed line) the wavefunction has a node at the Condon point ($R_C\approx$ 160\anul) and the overlap between the two channels is small, whereas for the second  value of the accumulated phase (dashed-dotted line) the overlap is maximal. \label{fg:FCfactor}}
\end{figure}

In Fig~\ref{simplemodel} we have plotted the resonance position $E_r$, the width $\Gamma$ and the phase difference $\phi$ as a function of the accumulated phase $\phi_s$ of the ``S+S'' potential at 20\anul, which we obtained by fitting the calculated cross sections using Eq.~\ref{BreitWigner}. If we change the phase of the ``S+S'' potential, the Franck-Condon overlap between the ``S+S'' and ``S+P'' wavefunction is changed (see Fig.~\ref{fg:FCfactor}). If the ``S+S'' wavefunction has a node at the Condon point, the overlap between the two wavefunctions is small. This leads in Fig.~\ref{simplemodel}b to a very narrow resonance for $\phi_s$ = 0.9$\pi$, whereas for $\phi_s = 0.45\pi$ the overlap is maximal and thus the width is the largest. The coupling not only leads to a broadening of the resonance profile, it also leads to a shift of the resonance position (see Fig.~\ref{simplemodel}a). This shift is called the light shift, since it is induced by the coupling of two potentials by the laser light. The sign of the light  shift depends on the fact, whether the overlap is stronger for distances smaller or larger than the Condon radius $R_C$. For smaller distances the ``S+P'' potential energy is lower than the ``S+S'' potential energy and thus the coupling between the two potentials pushes the ``S+P'' potential down in energy, which leads to a negative light shift. Conversely, if the overlap is larger at distances larger than $R_C$, this leads to a positive light shift. Surprisingly, if the overlap between the two wavefunctions is maximal and the coupling is the strongest ($\phi_s=0.45\pi$), the light shift becomes zero. This is caused by the fact that the ``S+S'' wavefunction peaks at the Condon point and the ``contributions'' to the light shift for smaller and larger distances cancel. Finally, we see in Fig~\ref{simplemodel}c that the phase difference $\phi$ is directly proportional to the accumulated phase $\phi_s$, which is to be expected in this simple model.

\subsection{Two channels with ionization}

\begin{figure}
    \centerline{\includegraphics[width=0.45\textwidth]{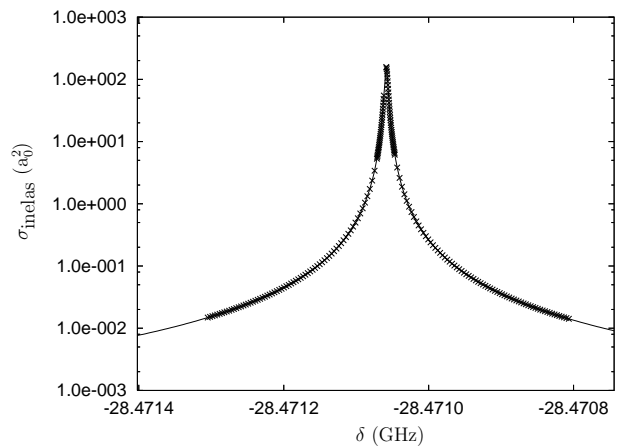} }
    \caption{Lorentz profile on a logarithmic scale in the inelastic cross section as a function of $\delta$. The crosses are numerical data and the line is a Lorentz profile with $E_r$ = $-$28.47106 GHz, $\Gamma$ = 4.770 kHz and $A$ = \EE{1.583}{2} \anul${}^2$. }\label{2dLorentz}
\end{figure}

We extend our model by adding an ionization process in the inner region of the excited-state potential and considering also the inelastic cross section. An example of a resonance in the inelastic cross section is shown in Fig.~\ref{2dLorentz}. Exactly as the elastic cross section, it contains resonances from bound states in the excited-state potential, which take the form of Lorentz profiles due to the absence of a direct ionization signal, given by
\begin{equation}
 \sigma_{\rm inelas} =   \frac{A \Gamma^2/4}{(\delta-E-E_r)^2+  \Gamma^2/4},
 \label{Lorentz}
\end{equation}
where $A$ is the height of the Lorentz peak, $\Gamma$ is its width
and $E_r$ is its position.

\begin{figure}
    \centerline{\includegraphics[width=0.45\textwidth]{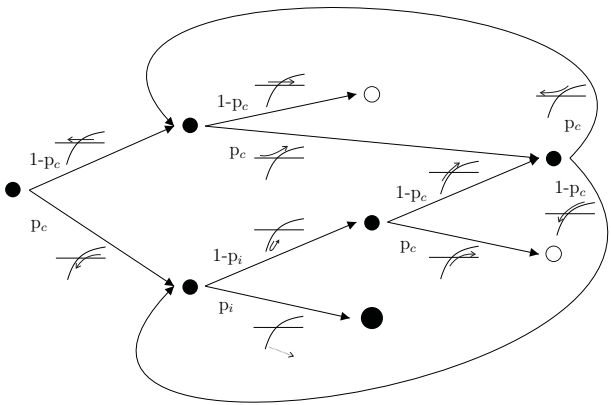}}
    \caption{Flux diagram for the two-channel problem with ionization. At each point (small circles) the system has a certain probability to make a transition at the crossing. In the end the system either scatters elastically (open circles) or inelastically (filled circles). }\label{fg:flux}
\end{figure}

For comparison, we develop a semi-classical picture, in which a metastable helium atom comes in on the ``S+S'' potential. At the Condon point, we denote with $p_c$ the probability that the atom follows the adiabatic route to the ``S+P'' potential. The probability that the atom ionizes in the inner region of the ``S+P'' potential is designated with $p_i$. From the inner regions of the two potentials, and from the outer region of ``S+P'', the atom, if not ionized, is assumed to reflect back to the Condon point.  In Fig.~\ref{fg:flux} a flux diagram is drawn for this semi-classical situation. It contains three simple loops (\idest, loops without external branches) that can be integrated out. The total probability $P_I$ for the atom to ionize is then found to be
\begin{equation}
P_I = p_c p_i \frac{1 + {(1-p_c)^2}/{(1 - p_c^2)}}{1 - (1 - p_i) {(1 - p_c)^2}/{(1 - p_c^2)}} = \frac{2p_c p_i}{p_c(2 - p_i) + p_i}. \label{LandauZener2D}
\end{equation}
In the middle part of this equation the three geometric series for the three loops can still be recognized. We can calculate the crossing probability p$_c$ using the Landau-Zener formula
\begin{equation}
p_c=1-e^{-\pi \Lambda}, \qquad \Lambda=\frac{\Omega^2}{2\alpha v_c},
\end{equation}
where $\Omega$ is the Rabi frequency between the ground and excited state, $\alpha$ is the difference in derivatives of the potentials at the Condon point, and $v_c$ is the velocity of the incoming atom at the Condon point. For a kinetic energy $E$ of 1.9 mK, a Rabi frequency of 66 kHz and a detuning of $\delta = -28.4$ GHz, we find $p_c$ = \EE{8.84}{-10}.

One should note that in such a semi-classical treatment of the problem we can no longer treat the dependence in  $\delta$ and only the total ionization probability is calculated. Quantum mechanically, however, we calculate cross sections that are  a function of $\delta$. As a measure of the total ionization probability, we have therefore taken the total area underneath one peak of the inelastic cross section. Although the integrated inelastic cross section obtained in this way is proportional to the total ionization probability, we lack a method to map the quantum-mechanical results onto the semi-classical picture. Therefore we can only compare the two methods with respect to their qualitative behavior as a function of the ionization probability $p_i$, but not with respect to the absolute total ionization probability $P_I$ they give rise to.

\begin{figure}
    \centerline{\includegraphics[width=0.45\textwidth]{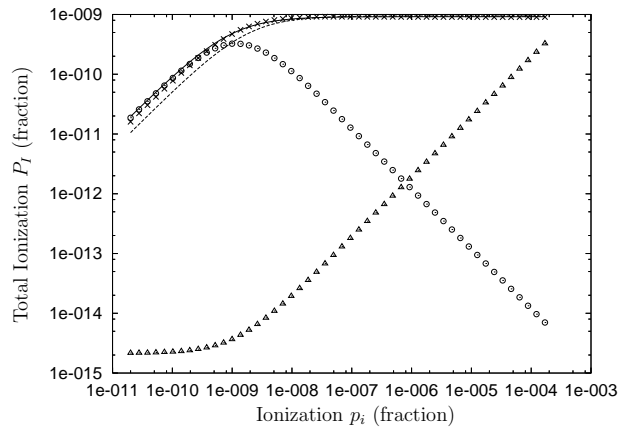}}
    \caption{Comparison between semi-classical and numerically calculated values for the total ionization probability $P_I$ as a function of the ionization probability $p_i$ in the excited-state potential. The crosses are the scaled numerical data, the solid line is Eq.~\ref{LandauZener2D} with a fitted value for $p_c$ (non-scaled), and the dashed line is the same equation with the Landau-Zener value for $p_c$ and scaled. The triangles are numerically calculated widths $\Gamma$ of the resonances, and the squares are their heights, both in arbitrary units.}\label{2dfitPicture}
\end{figure}

In Fig.~\ref{2dfitPicture} we plotted the results of both calculations as a function of the probability $p_i$. We normalized for large $p_i$ the total amplitude of the numerical calculation to the total ionization probability calculated semi-classically. If we allow $p_c$ to be an adjustable parameter, the best fit is for $p_c$ = \EE{4.59}{-10}, which is about half the Landau-Zener value. Note, that the Landau-Zener formula describes a single crossing, whereas in the current situation the atoms vibrates back and forth over the crossing leading to interference effects. From the figure  it can be seen that the widths of the resonances are increasing as a function of $p_i$, but their heights reach a maximum around the point where $P_I$ reaches its asymptote.

\subsection{Three channels with ionization}

\begin{figure}
    \centerline {\includegraphics[width=0.45\textwidth]{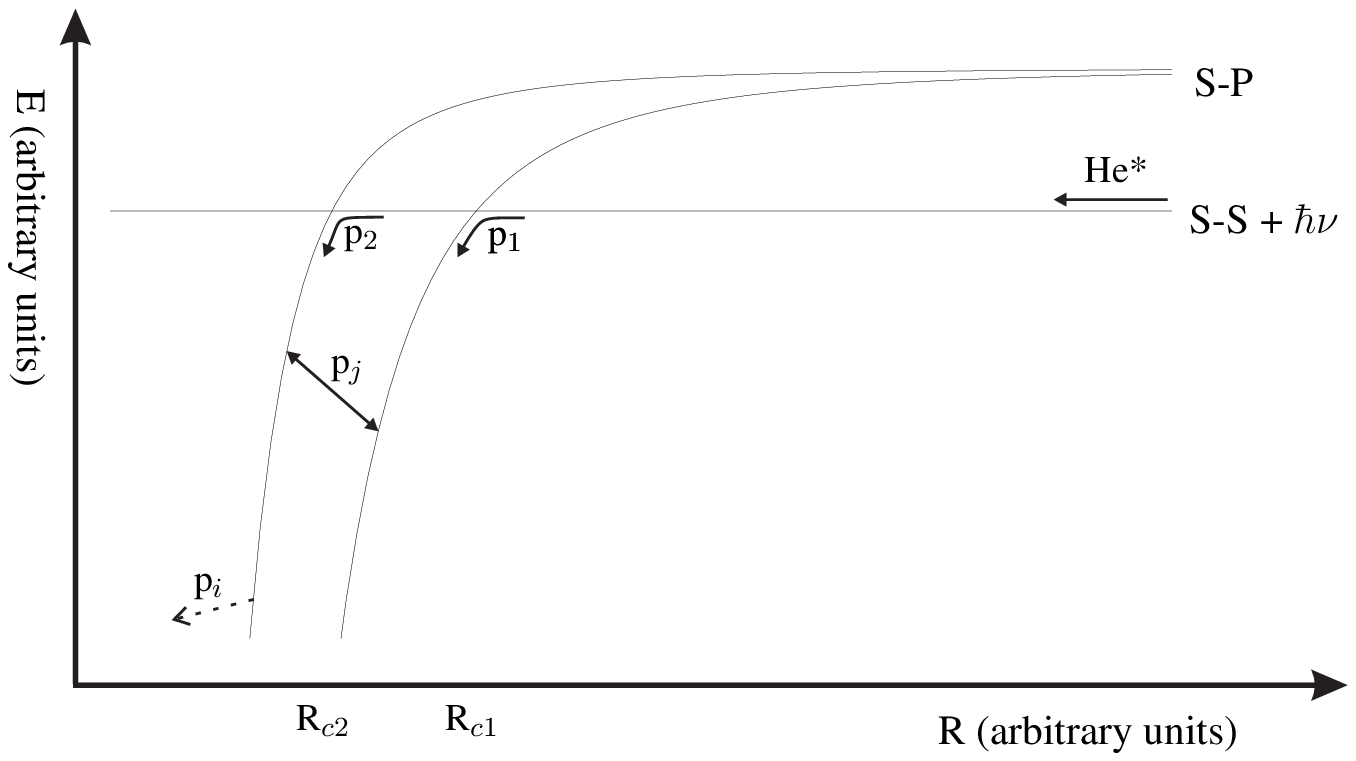}}
    \caption{The model system with two excited-state potentials. Only the ``S+P1'' potential is ionizing with probability $p_i$.} \label{3dsemiclassical}
\end{figure}

Next we consider the case where we have two excited-state potentials. We take both excited potentials to belong to the same ``S+P'' asymptote, but have slightly different $C_3$ values. These potentials will be denoted ``S+P1'' ($C_3=-9$) and ``S+P2'' ($C_3=-10$). Only the ``S+P1'' potential leads to  Penning ionization at short internuclear distance. The potentials are coupled by a rotational coupling with a probability $p_j$, which acts over a large distance. This system is depicted in Fig.~\ref{3dsemiclassical}. Since the effective coupling between two potentials is proportional to the wavefunction overlap and the wavefunctions of the closed channels have a maximum close to their Condon points, we expect the rotational coupling to be most effective before the first Condon point.  For the semi-classical treatment of the total ionization probability $P_I$, we assume that the coupling only takes place at this point, and that the atom can make a transition from one ``S+P'' potential to another both on the way in and on the way out. Furthermore, in the Landau-Zener model the transfer probability depends quadratically on the off-diagonal term in the Hamiltonian. 

\begin{figure}
    \centerline{\includegraphics[width=0.45\textwidth]{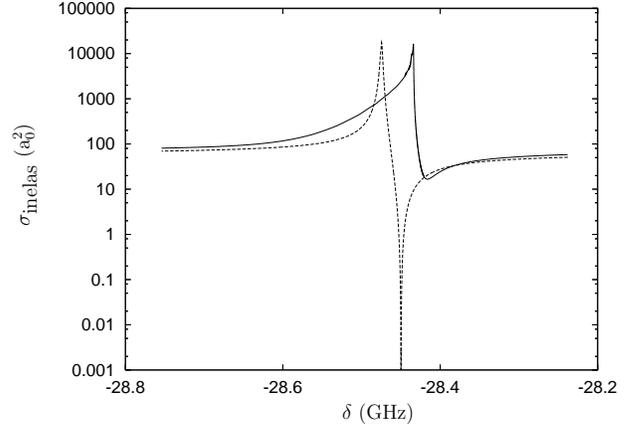}}
    \caption{Resonances in case of a single kinetic energy (dashed line) and a thermal distribution (solid line). The  Rabi frequency is $\Omega$ = 6.6 MHz for both couplings to the excited state, an ionization probability $p_i$ = 0.80 in ``S+P1'' and an energy of $E$ = 2.0 mK. For the thermal averaging the temperature is $E/k_B$.}\label{3D-TEffects}
\end{figure}

In Fig.~\ref{3D-TEffects}  we consider the resonance at $\delta = -28.4$ GHz, where we assume that the two states are coupled to the ground state with the same Rabi frequency. Since the Rabi frequency is large, the resonances from the ``S+P1'' potential are so wide that they overlap to a large extent. In contrast, bound states in the non-ionizing ``S+P2'' potential will give rise to much sharper peaks in the inelastic cross section, because only the part of the wavefunction coupled to the ionizing ``S+P1'' potential can lead to ionization. The inelastic cross section thus contains a background part caused by direct ionization in the ``S+P1'' channel, which interferes with resonances from the bound states in the ``S+P2'' potential. This leads to a Breit-Wigner profile with a non-constant background. The result of this interference can be quite unpredictable and instead of fitting the profiles we choose to numerically integrate the area underneath the resonance as a measure of the ionization probability.

To solve the semi-classical problem, we draw a flux diagram containing all the possible paths of the system. For every intersection in this diagram we write down an equation, and combine these equations into a set of coupled linear equations that is solved analytically. The solution for the total probability $P_I$ is given by
\begin{subequations}
\label{LandauZener3D}
\begin{align}
P_I & =\frac{2p_i\left((1-p_1)p_1p_2+x(p_j-p_j^2)\right)}{p_x+p_y+p_z}, \\
p_x & =p_1\left(p_2[2-2p_1(1-p_i)-p_i]+p_i-p_1p_i\right),\\
p_y & =\left(2x+\left(1-p_2-p_1[4-6p_2-p_1(3-7p_2)]\right)p_i\right)p_j,\\
p_z & =-\left(2x+(1-2p_1)[1-2p_2-p_1(1-3p_2)]p_i\right)p_j^2,\\
x & = p_2+(1-p_1)p_1(1-5p_2).
\end{align}
\end{subequations}
For the parameters used in Fig.~\ref{3D-TEffects} the Landau-Zener probabilities are given by $p_1$ = \EE{8.84}{-6} and $p_2$ = \EE{8.54}{-6}, where the small difference stems from the fact that the ``S+P1'' potential is less steep at its Condon point than the ``S+P2'' potential.

\begin{figure}
    \centerline{\includegraphics[width=0.45\textwidth]{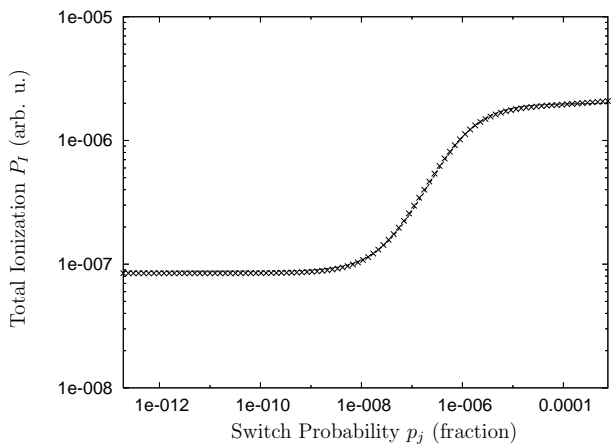}}
    \caption{Numerical data for the three-channel system as a function of the rotational probability $p_j$ (crosses) and the semiclassical formula of Eq.~\ref{LandauZener3D} fitted to the data (line), where we fit the semiclassical parameters $p_1$, $p_2$  and  the coupling strength between $p_j$ and the square of the off-diagonal matrix element. } \label{3dfitPicture}
\end{figure}

In Fig.~\ref{3dfitPicture} we compare the semi-classical and numerical calculations as a function of the rotational coupling $p_j$. The value of $P_I$ for large $p_j$ is determined by $p_1$, whereas for small $p_j$ it is determined by $p_2$. To get good agreement between the numerical analysis and the semi-classical description we have adjusted the parameters $p_1$ and $p_2$ and find $p_1$ = \EE{9.12}{-7} and $p_2$ = \EE{3.60}{-7}. Note that the relatively large difference between these two parameters, which are nearly identical in the Landau-Zener description (see above), reflects the fact that for these two states the Franck-Condon overlap with the ``ground'' state at the Condon point can be very different and thus that the behavior is dominated by Franck-Condon factors. Although the factors $p_1$ and $p_2$ calculated this way do not each have physical significance, their  ratio $p_1/p_2$ has. Numerically, we calculate this ratio to be 2.53, whereas Landau-Zener description yields 1.04. Clearly, we have to resort to quantum mechanical methods to solve this model, since the Landau-Zener formula assumes isolated avoided crossings, whereas we have two crossings lying close to each other.  From Fig.~\ref{3dfitPicture} it is clear that with the fitted parameters $p_1$, $p_2$ and $p_j$ the agreement between our numerical calculations and the semiclassical result is very good.

\subsection{Thermal distribution\label{sc:thermal}}

In our experiment, the interacting atoms do not all have the same relative energy, but instead have a thermal distribution determined by the temperature of the sample. We investigate the effect of the thermal distribution by integrating the profiles over a Maxwell-Boltzmann thermal distribution. The effect of the thermal averaging depends on the width of a single resonance compared to the Doppler shifts. In Fig.~\ref{3D-TEffects} we show a plot of the resonance at $\delta = -28.4$ GHz in the inelastic cross section for a relative energy $E=k_B T$ and for a thermal distribution. In the latter the dip has largely disappeared.  This is as expected, as the Doppler width and the resonance width are about the same size here ($\sim$\EE{6}{-9} E$_h$).

\section{Comparison between theory and experiment\label{sc:multi}}

Now we return to the numerical treatment of the experiment. As is clear from the previous section, the location of resonances in the inelastic cross section depends on the accumulated phases in the non-ionizing (\oneu\ and \twou) channels. In order to determine the accumulated phases, we first have to identify the  resonances belonging to one particular potential.  In a combined effort between our group and the group at ENS in Paris~\cite{ENSandAOpaper} this has been done for this system. This effort is based on the fact that all the resonances in one potential have (at least approximately) the same accumulated phase. So by calculating the accumulated phases needed to reproduce all resonances, one can group the resonances together.

As mentioned before, we are left with two sets of resonances; one for the \oneu\ potential and one for the \twou\ potential. Table \ref{resonancesTable} is a list of the resonances in the two channels. For both of these potentials we calculate the bound states as a function of the accumulated phase in a simple one-channel calculation. Selecting the accumulated phase that correctly reproduces one resonance of that potential indeed reproduces all the other resonances of that channel at the correct positions. The phases calculated using this procedure are 0.368$\pi$  for the \oneu\ potential and 0.429$\pi$  for the \twou\ potential. These turn out not to be the entirely correct phases to use in the multi-channel calculations, especially so for the \twou\ potential. In general, in calculations performed for the simple models of the previous section, accumulated phases for one and multi-channel calculations matched as expected. That this is not the case here is due to the strong coupling between the excited-state potentials, which causes the resonances to shift. This explanation is supported by model calculations for the three-channel model, in which the resonances are observed to shift for strong rotational coupling between the excited-state channels. Through fitting with experiment we find the accumulated phases to be 0.363$\pi$ for the \oneu\ potential and 0.445$\pi$ for the \twou\ potential.  This means that the rotational coupling has an effect on the exact location of the resonances. The accumulated phases of the two ionizing excited-state potentials have no influence on the spectrum and we simply take these phases equal to zero.

\begin{table}
    \center{\begin{tabular}{cc} \hline
        \oneu & \twou \\
        $-\delta$ (GHz)&$-\delta$ (GHz) \\ \hline
        11.10 & 13.57 \\
        7.01 & 8.94 \\
        4.26 & 5.64 \\
        2.42 & 3.38 \\
        1.21 & 1.88 \\
         & 0.98 \\
         & 0.46 \\
         & 0.19 \\ \hline
        \end{tabular}}
    \caption{Part of the set of resonances ($J=3$) in the two non-ionizing potentials being considered. } \label{resonancesTable}
\end{table}

\subsection{Accumulated phase ground state}

In the previous section we discussed the selection of the appropriate accumulated phases for the excited states. However, also in the ground-state channel we need to select the correct accumulated phase. This phase has a very large influence on the $s$-wave scattering length as this scattering length depends directly on the final phase shift of the wave function. Recently Moal \etal~\cite{Moal} measured it very accurately to  141.96 $\pm$ 0.09 \anul. We choose the accumulated phase for the ground state to match this value using~\cite{Joachain}
\begin{equation}
a = - \lim_{k \rightarrow 0} \frac{\delta_0}{k},
\end{equation}
where $a$ is the $s$-wave scattering length. We use a low energy of 19 $\mu$K and a low laser intensity of $s_0$ = 10. That way we find the accumulated phase to equal 0.939$\pi$.


\subsection{Spectrum}

\begin{figure}
    \includegraphics[width=0.45\textwidth]{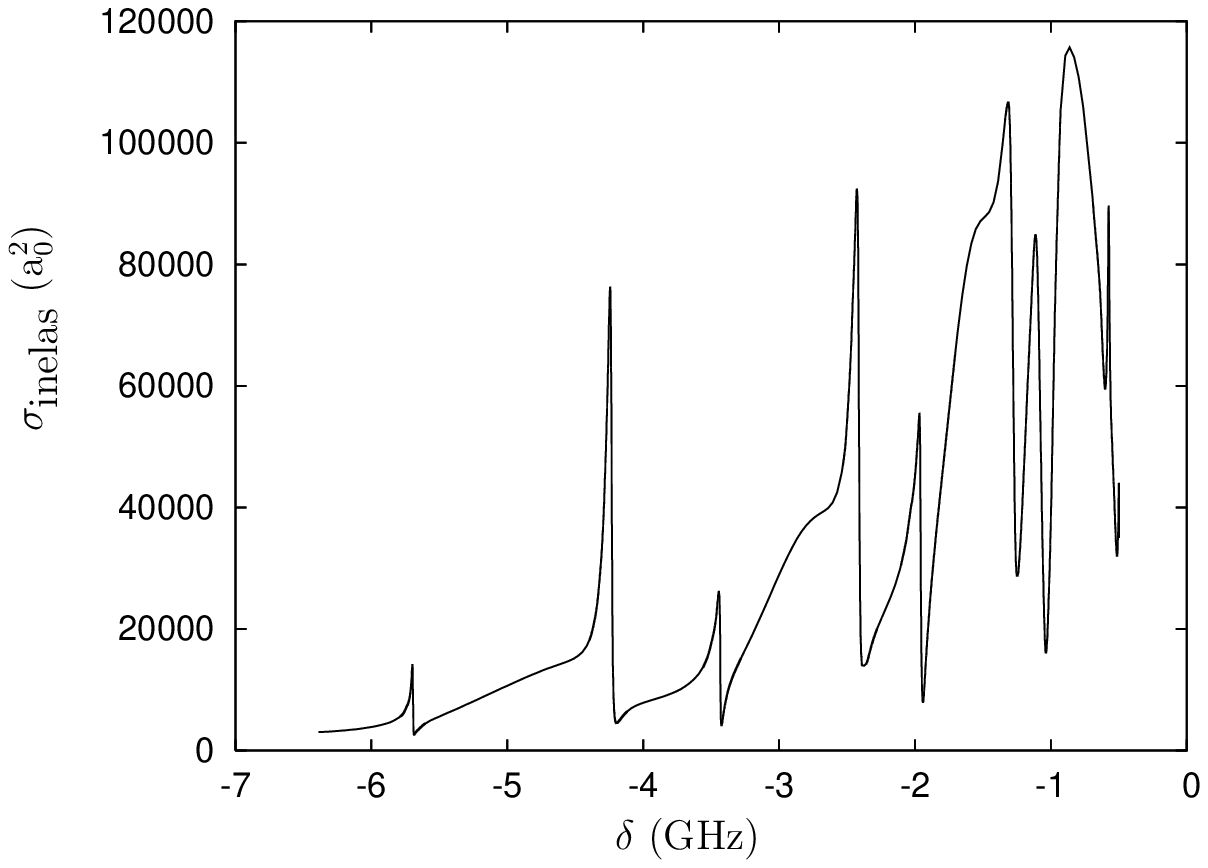}
    \centerline{(a)} \\
    \includegraphics[width=0.45\textwidth]{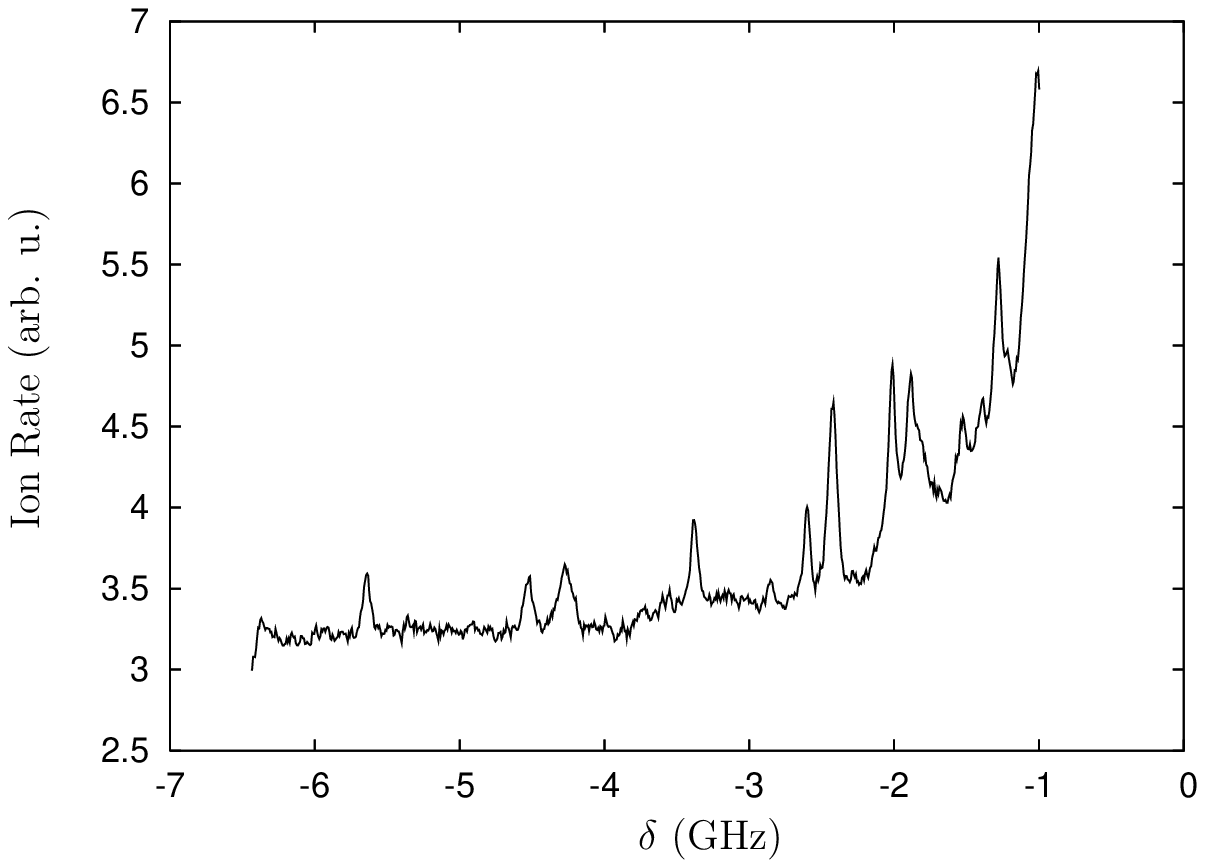}
    \centerline{(b)} \\
    \caption{Figure (a) is a numerical spectrum for $T$ = 1.9 mK and $s_0$ = \EE{1}{4}. Figure (b) is an experimental spectrum for $T$ = 1.9 mK and $s_0$ a few times \ee{4}.}\label{spectrum}
\end{figure}

To obtain a numerical spectrum that we can compare with experimental results, we use the accumulated phases from the previous sections and we integrate over a thermal distribution as explained in Sec.~\ref{sc:thermal}. Figure \ref{spectrum}a is calculated for a temperature of 1.9 mK and a laser saturation parameter of $s_0$ = \EE{1}{4}. The former is in accordance with experiment, the latter is low in comparison with experiment. A comparison with the experimental spectrum that is reproduced in Fig.~\ref{spectrum}b shows that we successfully reproduce the resonances at the correct positions. Both spectra contain resonances that have widths in between roughly 100 and 200 MHz. However, the thermal effects do not cause enough smearing  of the resonances to make the resonances symmetric in our calculations, whereas they look completely symmetric in the experimental data. The theoretical spectrum contains also a much higher peak-to-background ratio than the experimental data.  Note that also the so-called {\it optical collisions}, \idest, the increase of the background level as we approach zero detuning, is present in our calculations. This is because for small detunings many bound states are available in the excited-state channels. Therefore a very large number of resonances of those channels contribute to an increased ionization rate. Note, that for small detunings our assumption that we only have $s$-wave scattering breaks down.

\subsection{Line widths}

\begin{figure}
    \centerline{\includegraphics[width=0.45\textwidth]{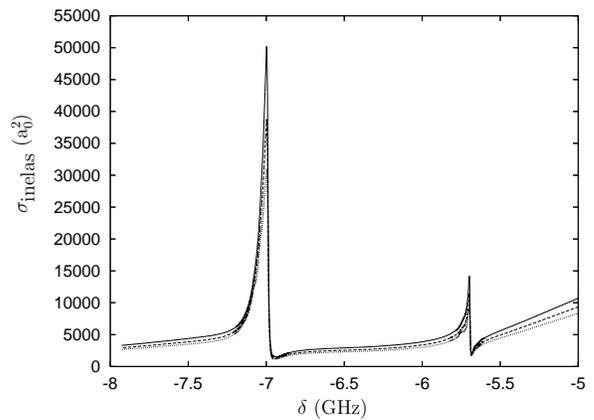}}
    \caption{Two resonances at different temperatures. Solid line is 1.9 mK, dashed line is 2.3 mK and dotted line is 2.7 mK.} \label{tempdependence}
\end{figure}

To study in more detail the effects that determine the line widths of the resonances in the numerical spectrum, we perform the same calculation as in the previous paragraph, but now for different temperatures. The results can be found in Fig.~\ref{tempdependence}. The figure shows that a higher temperature will lead to a slightly narrower resonance. This is counterintuitive, since one would expect that a higher temperature leads to a larger Doppler broadening. This is also observed for the model systems of Sec.~\ref{sc:model}. In this case the width of a resonance seems to be determined by other effects. This is in accordance with what is observed in the experiment. 

\begin{figure}
    \centerline{\includegraphics[width=0.45\textwidth]{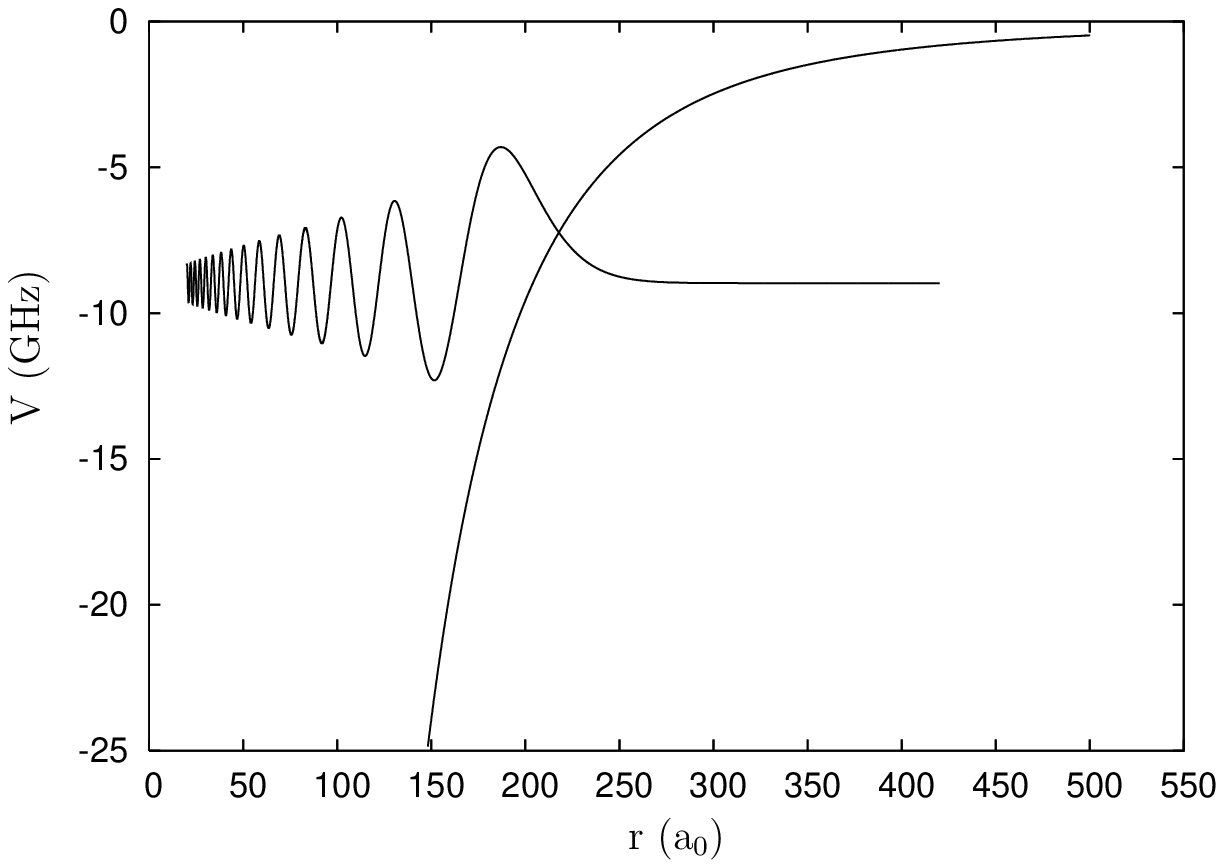}}
    \caption{The wavefunction of the bound state at $\delta=-$8.94 GHz depicted in its \twou\ potential. }  \label{FranckCondon}
\end{figure}

One possible suggestion is that the line widths are determined by the last oscillation of the bound state at the Condon point. This final part of the wavefunction is much larger than the remainder of the wavefunction, and thus could therefore dominate the Franck-Condon factor. Its projection on the potential gives us the distance over which photoassociation can occur, and thereby also the range of detunings that contribute to the photoassociation process. In Fig.~\ref{FranckCondon} we show that at the temperature used in the experiment the contribution to the Franck-Condon factor of the other oscillations is significant. Furthermore the last oscillation covers a range of the order of 50\anul, leading to a detuning range of 10 GHz. This is much wider than the resonance widths observed in the experiment, and we conclude that this last oscillation is not the limiting factor for the resonances studied.

\section{Conclusions}

\hyphenation{asymp-tote}
In this paper we have studied the role of Penning ionization for the observation of photoassociation in the He$_2$(\tripS-\tripS) system. We have shown that the resonances in experimental ionization rates are due to bound states in the long-range potentials near the \tripS-\tripP\ asymptote that are reached through photoassociation. We can describe this process with the multi-channel model discussed in Sec~\ref{sc:multi}, where the coupling between bound, ionizing channels and bound, non-ionizing channels leads to well-resolved peaks in the ionization spectrum. Although the line shape is not completely understood from the model, the location of the lines is in good agreement with the predictions of the model. The study shows that photoassociation spectroscopy is a powerful technique at ultracold temperatures even in systems where inelastic processes play a major role.

\section{Acknowledgments}

The work of DN is supported by the EU research training network ``Cold molecules'' (COMOL), under the contract number HPRN-2002-00290.


\end{document}